\documentclass[a4paper]{jpconf}

\usepackage[latin1]{inputenc}
\usepackage{graphicx}

\usepackage{hyperref}
\hypersetup{
		colorlinks,
		citecolor=black,
		filecolor=black,
		linkcolor=black,
		urlcolor=black
	}
\usepackage{xcolor}

\usepackage{cancel}

\begin{document}

\title{Chemical freeze-out of light nuclei in high energy nuclear collisions and resolution of the hyper-triton chemical freeze-out puzzle}

\author{K. A. Bugaev$^{1, 2}$,
O. V.  Vitiuk$^{2, 3}$,
B. E. Grinyuk$^{1}$, 
N. S. Yakovenko$^{2}$,
E. S. Zherebtsova$^{4, 5}$,
V. V. Sagun$^{1,6}$, 
O. I. Ivanytskyi$^{1,6}$,  
D. O. Savchenko$^{1}$,  
L. V. Bravina$^3$,
D. B. Blaschke$^{4,7,8}$, 
G. R.  Farrar$^{9}$,
S. Kabana$^{10}$, 
S. V. Kuleshov$^{11}$,
E. G. Nikonov$^{12}$,
A. V. Taranenko$^6$,
E. E. Zabrodin$^{3, 13}$ 
and 
G. M. Zinovjev$^1$ }

\address{$^1$ Bogolyubov Institute for Theoretical Physics, Metrologichna str. 14-B,  03680 Kyiv, Ukraine}
\address{$^2$  Department of Physics, Taras Shevchenko National University of Kyiv, 03022 Kyiv, Ukraine}
\address{$^3$ Department of Physics, University of Oslo, PB 1048 Blindern, 
N-0316 Oslo, Norway}
%%%
\address{$^4$ National Research Nuclear University (MEPhI), Kashirskoe Shosse 31, 115409 Moscow, Russia}
\address{$^5$ Institute for Nuclear Research, Russian Academy of Science, 108840 Moscow, Russia}
\address{$^6$ CFisUC, Department of Physics, University of Coimbra, 3004-516 Coimbra, Portugal}
%%%
\address{$^7$   Institute of Theoretical Physics, University of Wroclaw, Max Born Pl. 9, 50-204 Wroclaw, Poland}
\address{$^8$ Bogoliubov Laboratory of Theoretical Physics, JINR, Joliot-Curie Str. 6, 141980 Dubna, Russia}
\address{$^9$ Department of Physics, New York University, 726 Broadway, 
    New York, NY 10003, USA}
\address{$^{10}$  Instituto de Alta Investigaci\'on, Universidad de Tarapac\'a, Casilla 7D, Arica, Chile}
\address{$^{11}$  Departamento de Ciencias �F\'{\i}sicas, Universidad Andres Bello, Sazi\'e 2212, 
%%Piso 7, 
Santiago, Chile}
\address{$^{12}$   Laboratory for Information Technologies, JINR, Joliot-Curie str. 6, 141980 Dubna, Russia}
\address{$^{13}$ Skobeltzyn Institute of Nuclear Physics, Moscow State University,
119899 Moscow, Russia}

\ead{Bugaev@th.physik.uni-frankfurt.de}

\begin{abstract}
We present a summary of the recent results obtained with the novel hadron resonance gas model with the multicomponent hard-core repulsion which is extended to describe the mixtures of hadrons and light (anti-, hyper-)nuclei. A very accurate description is obtained for the hadronic and  the light nuclei data measured by STAR at the collision energy  $\sqrt{s_{NN}} =200$ GeV and by ALICE at $\sqrt{s_{NN}} =2.76$ TeV.  The most striking result discussed here  is that for  the most probable   chemical freeze-out scenario for the STAR energy the found parameters allow us to  reproduce the values of the experimental ratios $S_3$ and $\overline{S}_3$ without fitting.
\end{abstract}

\section{Introduction}

The development of the  hadron resonance gas model  (HRGM) with the multicomponent hard-core repulsion between the constituents \cite{MHRGM0,MHRGM1,MHRGM2,MHRGM3,MHRGM4,Sagun14}, i.e. with several hard-core radii of hadrons,  converted the so-called thermal model into a powerful and convenient tool of the heavy ion physics phenomenology,  but  also it led to a few real breakthroughs in our understanding of the chemical freeze-out (CFO) process. Indeed, using just a few extra parameters  compared to the traditional HRGM \cite{Andronic:05}, which employs  a single  hard-core radius for baryons $R_b$ and the  one  $R_m$ for the mesons, it was possible to reach  an unprecedented accuracy in the description of  hadronic yields  measured in the central nuclear collisions  from the low AGS BNL collision energy ($\sqrt{s_{NN}} =2.7$ GeV) to the  highest RHIC one  ($\sqrt{s_{NN}} =200$ GeV)  with a quality  $\chi^2/dof \simeq 1.15$ \cite{MHRGM2,MHRGM3,MHRGM4} (if one includes into the fitting the hard-core radius of pions $R_\pi$ and kaons $R_K$) or with $\chi^2/dof \simeq 0.96$ \cite{Sagun14} (if one includes into the fitting  the hard-core radius of $\Lambda$-(anti-)hyperons $R_\Lambda$ in addition to  $R_\pi$ and $R_K$).

The high accuracy  achieved by the HRGM with multicomponent hard-core repulsion allowed us not only to elucidate the  characteristics  of the CFO of A+A collisions, but also to resolve several long-standing  problems of the CFO process \cite{MHRGM1,MHRGM2,MHRGM3,MHRGM4,Sagun14,GSA15,GSA16,GSA16b,Signals18,Signals19}: (i) in Refs. \cite{MHRGM1,MHRGM2,Sagun14} it was shown that  the so-called (anti-)$\Lambda$ puzzle \cite{Andronic:05}  is the result of oversimplifying assumptions; (ii) in Refs.  \cite{MHRGM1,MHRGM2} it was found a natural solution to the Strangeness Horn \cite{Horn} description puzzle which troubled the heavy ion community for a decade; (iii) the concept of separate CFOs of strange and non-strange hadrons was independently suggested in Refs. \cite{MHRGM3,Chatterjee1}. Moreover,  the high quality of data description allowed us to find out  several  new irregularities of thermodynamic quantities at the CFO which helped us  to formulate new and promising  signals of two phase transitions \cite{MHRGM4,GSA15,GSA16,GSA16b,Signals18,Signals19}  that  are expected to exist  in strongly interacting matter \cite{BugaevR1,BugaevR2,BugaevR3}.  

One should, however,  remember that the multicomponent versions of the HRGM \cite{MHRGM0,MHRGM1,MHRGM2,MHRGM3,MHRGM4,Sagun14}  based on the  popular Van der Waals (VdW) approximation to the hard-core repulsion, i.e. which use the classical second virial coefficients, are rather complicated and  take a lot of CPU time, since for  $N$ different hard-core radii  for each iteration of the experimental data  fitting  it is necessary  to solve the  system of $(N+1)$ transcendental equations containing  a few hundreds of double integrals. Hence, the application of the multicomponent HRGM based on VdW approximation to cases of $N \gg 1$ is somewhat   problematic \cite{IST2,IST3}. However, an entirely new and efficient  approach to deal with the multicomponent hard-core repulsion  in the grand canonical ensemble for  large values of $N$ was invented in Ref.  \cite{IST1}.  

This   novel approach is based on  the induced surface tension (IST)  concept \cite{IST1}. It  has two principal  advantages over the other multicomponent versions of the HRGM: (i) the number of equations which should be solved is two only and  does not depend on $N$, and  (ii) as shown in  \cite{IST2,IST3,QSTAT2019,Nazar2019} it allows one to go far beyond the usual  VdW approximation and to take into account not only  the second, but the third and even  the fourth virial coefficients of the classical hard spheres. In Refs. \cite{IST2,IST3} it was recently  shown that, in contrast to  the oversimplified version of the HRGM like the one used in \cite{KAB_PBM14}, there is  no proton yield puzzle neither at ALICE energy $\sqrt{s_{NN}} =2.76$ TeV, nor  at RHIC energies of collisions. 

Our next step was to extend  the IST equation of state (EoS) to the description of the mixtures of the hadrons with light nuclear clusters, i.e.  the  deuterons (d),  helium-3 ($^3$He),  helium-4 ($^4$He) and hyper-triton ($^3_\Lambda$H) and their antiparticles, and to apply the developed EoS to the simultaneous description  of   the  STAR  $\sqrt{s_{NN}} =200$ GeV data on the nuclear multiplicities \cite{STARA1,STARA2,STARA3}, the ALICE $\sqrt{s_{NN}} =2.76$ TeV data   on light nuclear cluster yields  \cite{KAB_Ref1a,KAB_Ref1b,KAB_Ref1c} and the hadronic multiplicities measured at these collision energies.   To our great surprise  not only the quantum, but also the classical  second virial coefficients of such nuclei and hadrons were never discussed in the literature.  Therefore, we had to resolve this problem first. Since the HRGM with the classical second virial coefficients of hadrons with the hard-core repulsion, i.e. with the excluded volumes,   is rather successful, we extended this approach  to the classical second virial coefficients of hadrons  and light nuclear clusters \cite{KAB_SepFO1,KAB_SepFO2,KAB_SepFO3,KAB_SepFO4}. 

In this work we summarize our  very recent results \cite{KAB_SepFO3,KAB_SepFO4} obtained on  the description of the STAR  $\sqrt{s_{NN}} =200$ GeV data on the nuclear multiplicities \cite{STARA1,STARA2,STARA3} and  the ALICE $\sqrt{s_{NN}} =2.76$ TeV data   on light nuclear cluster yields  \cite{KAB_Ref1a,KAB_Ref1b,KAB_Ref1c}, and discuss some findings  which were not reported previously, in particular, the problematic  hyper-triton  ratios (PHTR)    $S_3 = _\Lambda^3{H}/ ^3{He} \cdot p/\Lambda $  and $\overline S_3 $  measured  by the STAR  and ALICE Collaborations  which were  not described  until now either by the HRGM or by the coalescence model \cite{Rapp2018}. 

\section{HRGM  for the mixture of hadrons and light nuclear clusters}

The HRGM based on the IST EoS  has the following hard-core radii  of pions $R_{\pi}$=0.15 fm, kaons $R_{K}$=0.395 fm,  
$\Lambda$-(anti-)hyperons $R_{\Lambda}$=0.085 fm, other baryons $R_{b}$=0.365 fm and  other mesons $R_{m}$=0.42 fm \cite{IST2,IST3,Signals18} which only slightly differ from the our previous results found within the VdW approximation \cite{MHRGM3,Sagun14}. 
Since  all the details of the IST EoS and the fitting procedure of the hadronic  data are well documented in Refs. \cite{IST2,IST3,Signals18}, here we do not discuss them.

To  account  for the classical excluded volumes of light nuclear clusters and hadrons  we use  two approaches worked out in  \cite{KAB_SepFO1,KAB_SepFO2,KAB_SepFO3} with one exception, namely we consider the hyper-triton (HTR) differently  as it is suggested in \cite{KAB_SepFO4}. Both of  these approaches  employ  the  classical excluded volumes of light nuclei  of  $A \in \{2, 3, 4\}$ baryonic constituents   and hadron $h$ \cite{KAB_SepFO3,KAB_SepFO4}
\begin{eqnarray}\label{Eq1}
&& b_{Ah} = b_{hA} =  A \cdot \frac{2}{3}\pi (R_b+R_h)^3 , \qquad {\rm except~the~HTR}\,, \quad \\
 \label{Eq2}
&&  b_{HTRh} = b_{hHTR} =  2 \cdot  \frac{2}{3}\pi (R_b+R_h)^3 + \frac{2}{3}\pi (R_\Lambda+R_h)^3 ,  \qquad 
  {\rm for~the~HTR}\,.
\end{eqnarray}
The equations above  can be  found from the fact that all  light nuclear clusters analyzed here are roomy clusters. The mean distances among the baryons inside of such clusters  are rather large \cite{KAB_SepFO3,KAB_SepFO4} and, hence,   it is possible to  freely translate any  hadron $h$ with the hard-core radius $R_h$ around each  constituent of a nucleus  without touching  any other constituent  of   this  nucleus.  

The first approach is the IST$\Lambda$ EoS and it uses exactly the excluded volumes (\ref{Eq1}) and (\ref{Eq2}). It  is rigorously derived using a self-consistent treatment of classical excluded volumes of light nuclear clusters  and hadrons \cite{KAB_SepFO3} with the help of  the methods developed in \cite{QSTAT2019,Nazar2019}. In contrast, the IST EoS which employs  (\ref{Eq1}) for the HRT   is  called the  IST EoS.

The second approach is approximate and complementary to the exact  one. It is based on an approximate, but the rather accurate  treatment of the equivalent hard-core radius of roomy nuclear cluster and pions which are the dominating component of the HRG at the energy range of our interest. In the latter  approach  one can find an effective hard-core radius of nuclei  of $ A$ baryons  as $R_A \simeq A^{1/3}R_b$, since the  hard-core radius of pions is very small and, hence,  it generates a negligible correction to $R_A$  \cite{KAB_SepFO1,KAB_SepFO2,KAB_SepFO3}. Since the hard-core radius of light nuclear clusters defined in this way  is similar to the expression of the Bag Model  \cite{BMR}, it  is called the BMR EoS. A more accurate expression for  the HTR hard-core radius $R_{HTR} \simeq 2^{1/3} R_b$ is derived  in \cite{KAB_SepFO4} and such a model is called the BMR$\Lambda$ EoS. The main reason to compare the results of these two approaches is that, despite the difference in the equations, they should  reproduce   the data with the same quality by construction. Hence, finding the region of parameters which provide a similar quality of the data description one can remove the ambiguity in choosing the appropriate  CFO parameters by analyzing the wide and shallow  minima $\chi^2_A$ of light nuclei. 

\begin{figure}[th]
\centerline{
\includegraphics[width=0.5\columnwidth,height=63mm]{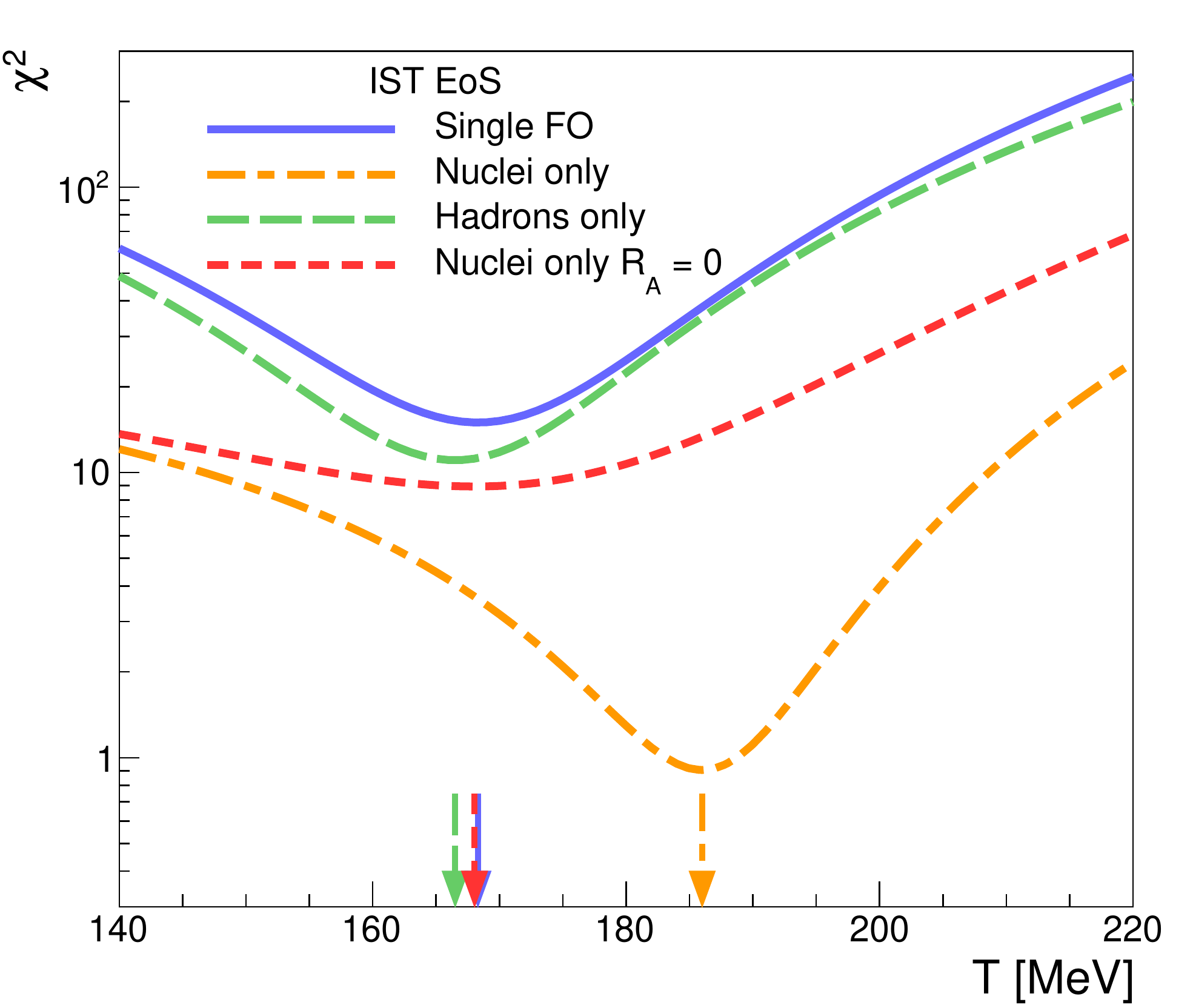}
	\hspace*{-1mm}
	\includegraphics[width=0.5\columnwidth,height=63mm]{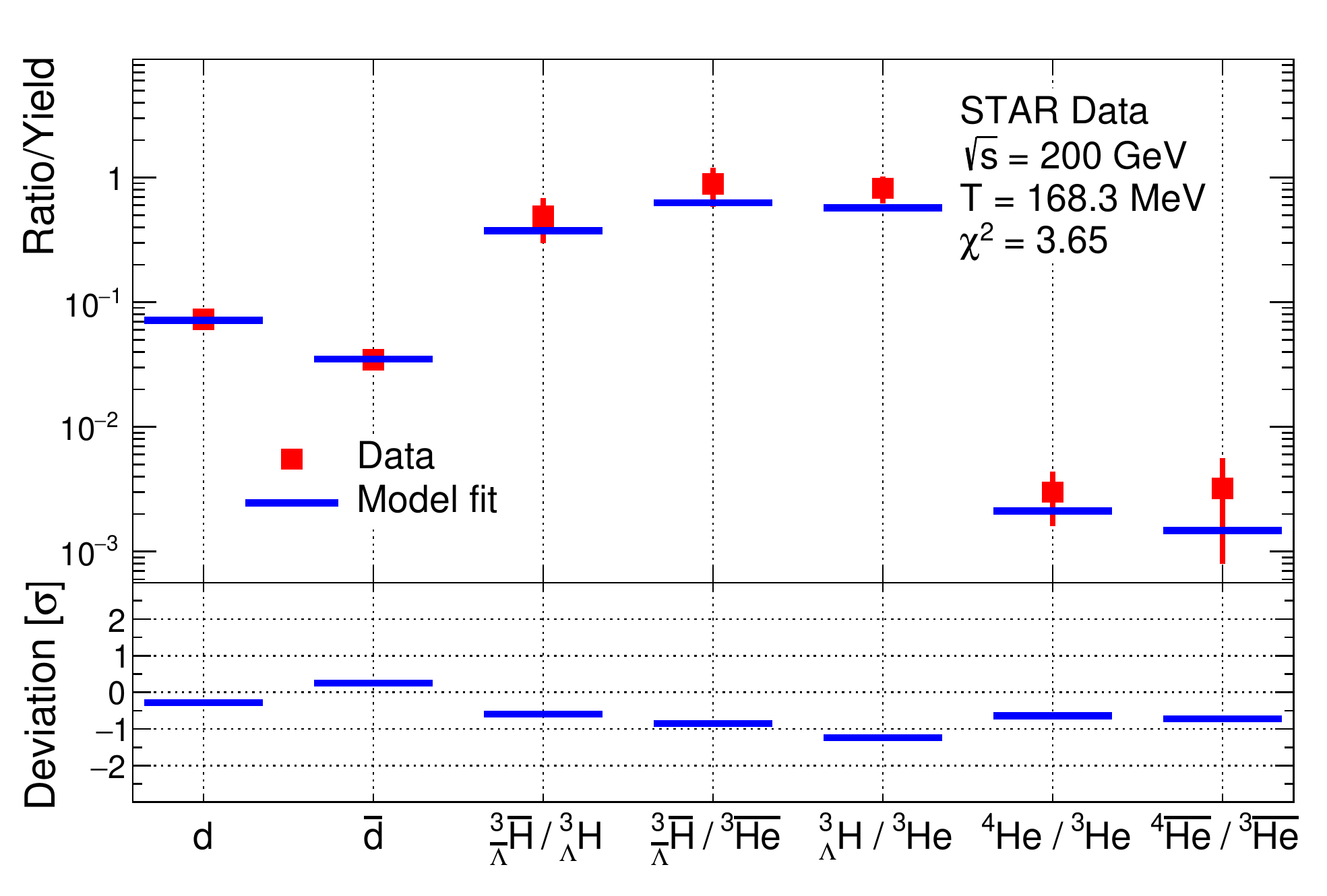}
}
		\caption{{\bf Left panel:}  Temperature dependence of $\chi^2_{tot}$, $\chi^2_{h}$ and $\chi^2_{A}$ for fit of the STAR data measured at $\sqrt{s_{NN}} = 200$ GeV found for the IST$\Lambda$ EoS.  
	{\bf Right panel:} The yields of nuclear clusters measured at $\sqrt{s_{NN}} = 200$ GeV by STAR vs. theoretical description with IST$\Lambda$ EoS. Insertion shows the deviation of theory from data in	the units of experimental error.   
}
	\label{KAB_Fig1}
\end{figure}
\begin{figure}[th]
\centerline{
\includegraphics[width=0.5\columnwidth,height=63mm]{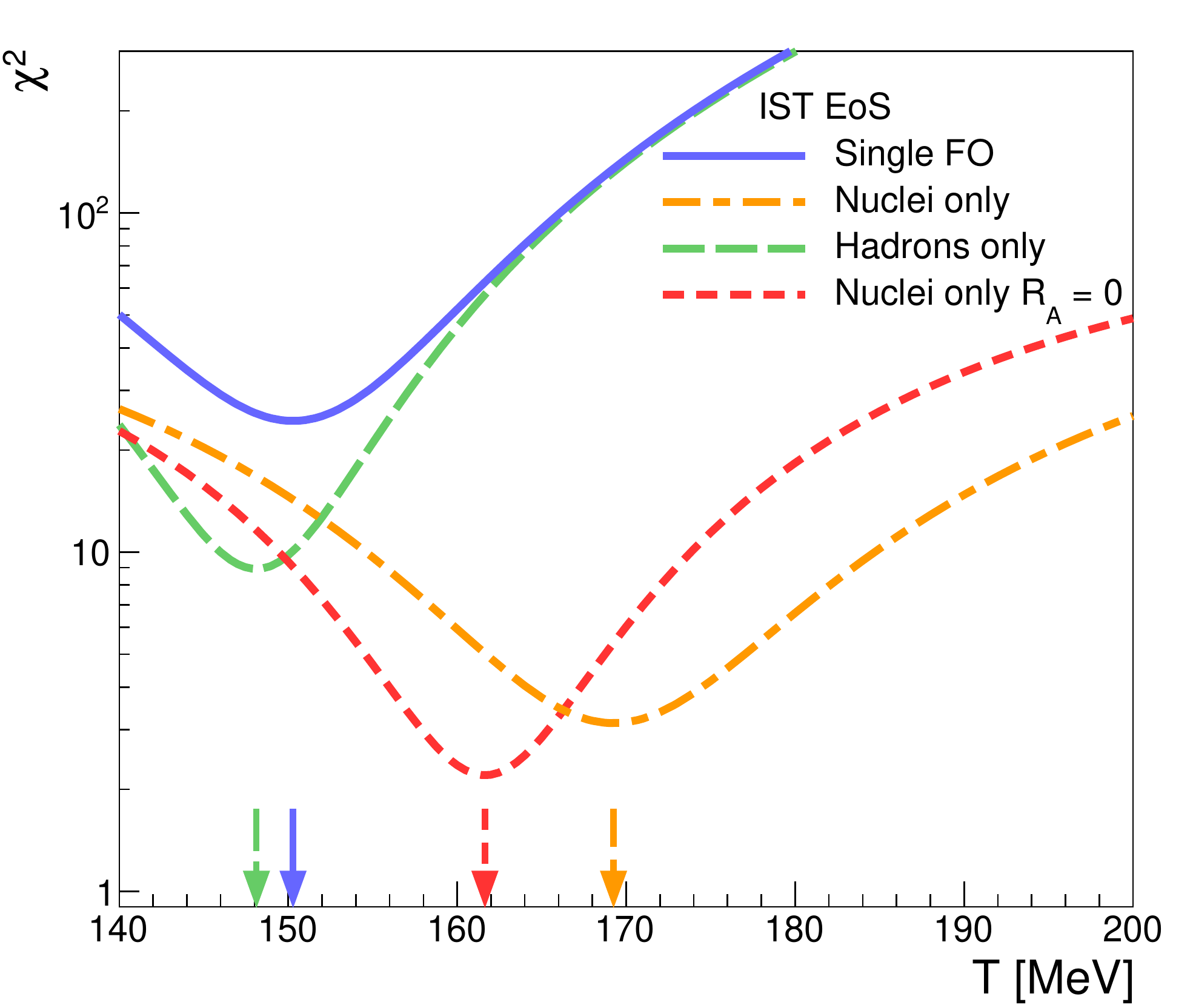}
	\hspace*{-1mm}
	\includegraphics[width=0.5\columnwidth,height=63mm]{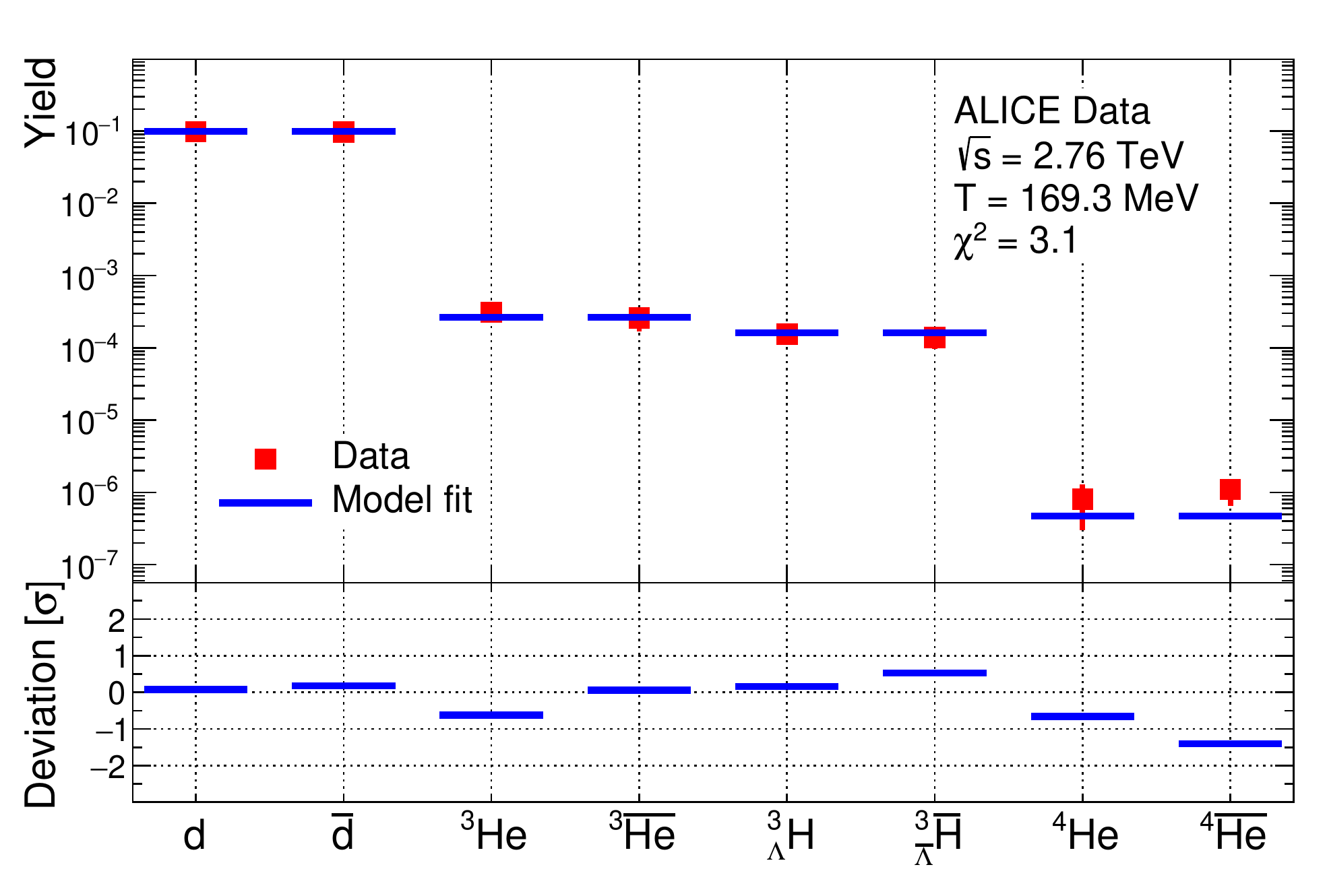}
}
		\caption{{\bf Left panel:}   Same as in Fig.~1, but for the  fit of the 
		ALICE data measured at $\sqrt{s_{NN}} = 2.76$ TeV.
		%%Temperature dependence of $\chi^2_{tot}$, $\chi^2_{h}$ and $\chi^2_{A}$ for fit of the 
		%%ALICE data measured at $\sqrt{s_{NN}} = 2760$ GeV found for the IST$\Lambda$ EoS.  
%%	{\bf Right panel:} The yields of nuclear clusters measured at $\sqrt{s_{NN}} = 2760$  GeV 
%%by ALICE  vs. theoretical description with IST$\Lambda$ EoS. Insertion shows the deviation of 
%%theory from data in	the units of experimental error.   
}
	\label{KAB_Fig2}
\end{figure}

Following our ideology  outlined in \cite{KAB_SepFO1},  we verify two different scenarios of the CFO of  nuclei clusters, namely a  single CFO together with the hadrons  and their  separate CFO from the hadrons. The major  reason for such an analysis is that the mechanisms of the hadron production and production of nuclei in collisions can be rather different.  One can clearly see from the left panels  of  Figs~\ref{KAB_Fig1} and \ref{KAB_Fig2} that  the  minimum of  the  light nuclear clusters $\chi^2_A (T_A)$ as a function of their  CFO temperature $T_A$ is located far away from  the minimum of   $\chi^2_h (T_h)$  of hadrons  as the function of  the hadronic CFO temperature $T_h$.  The total $\chi^2_{\rm tot}(V)$ is  defined as
\begin{eqnarray}
\label{Eq3}
%\begin{multlined}
\chi^2_{\rm tot}(T_h, T_A, V) &=& \chi^2_{R} + \chi^2_{Y}(V)
=  \sum_{ {k \neq l} \in R} \left[ \frac{{\cal R}_{kl}^{\rm theo}  - {\cal R}_{kl}^{\rm exp}}{\delta {\cal R}_{kl}^{\rm exp}}\right]^2  
% \nonumber \\ &&
\hspace{-2mm} + \hspace{-1mm}\sum_{k \in Y} \left[ \frac{\rho_k  V - N^{\rm exp}_k}{\delta N^{\rm exp}_k}\right]^2 , 
\end{eqnarray}
where $\chi^2_R$ and $\chi^2_Y$ denote, respectively, the mean squared deviation for the ratios and for the yields, while $V$ is the CFO volume of  nuclei and  $\rho_k$ is the particle number density of the $k$-th sort of particles.  A  combined fit of  particle  yields and ratios is  dictated by the available data and by numerical convenience. It is interesting that for the vanishing hard-core radii of all nuclei  the  minimum of $\chi^2_A (T)$ is close to the minimum of  $\chi^2_h (T)$ for the STAR data, but still it is far away for the ALICE one (see the short dashed curves in the left panels of Figs~\ref{KAB_Fig1} and \ref{KAB_Fig2}).

\begin{figure}[th]
\centerline{
\includegraphics[width=0.5\columnwidth]{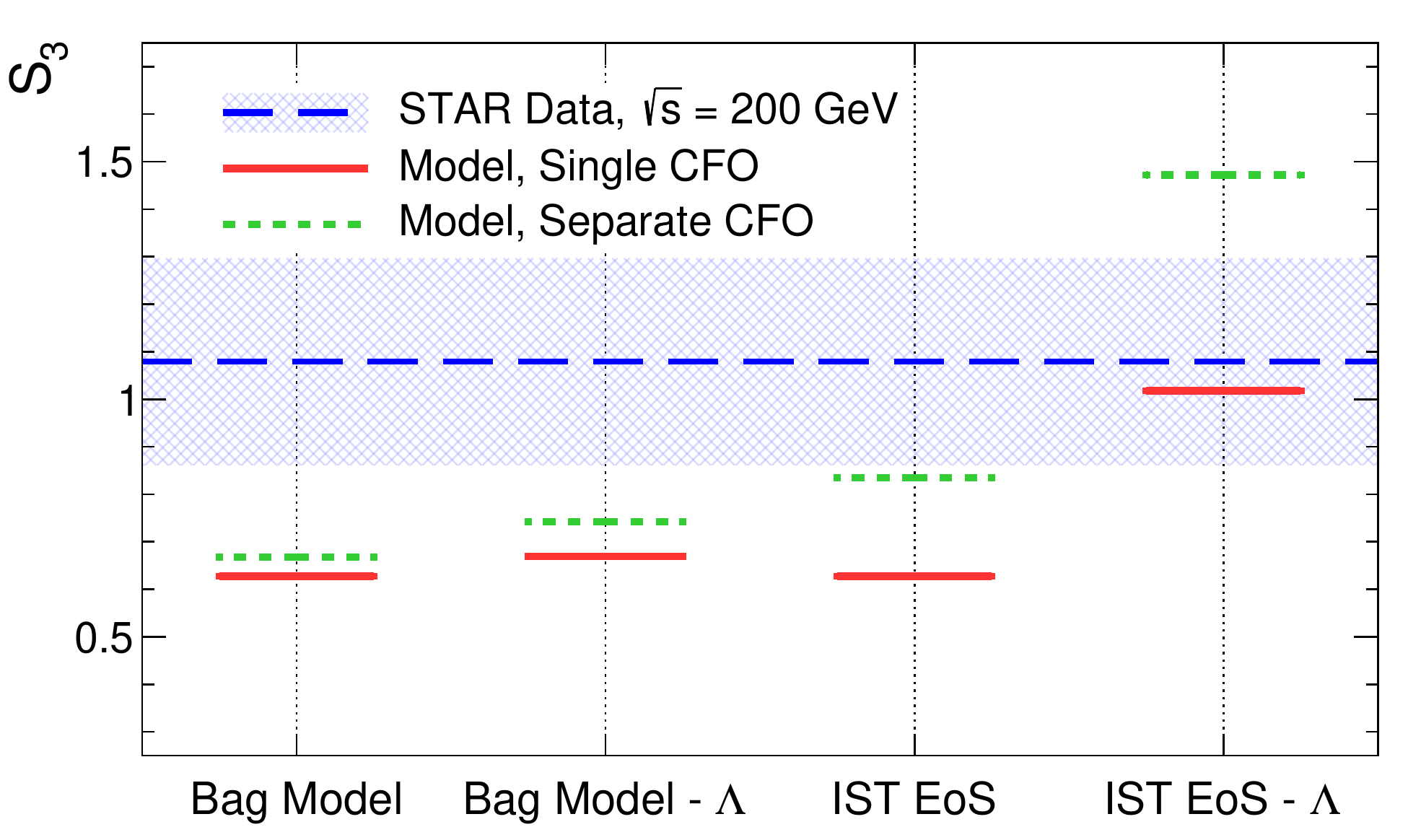}
	\hspace*{-1mm}
	\includegraphics[width=0.5\columnwidth]{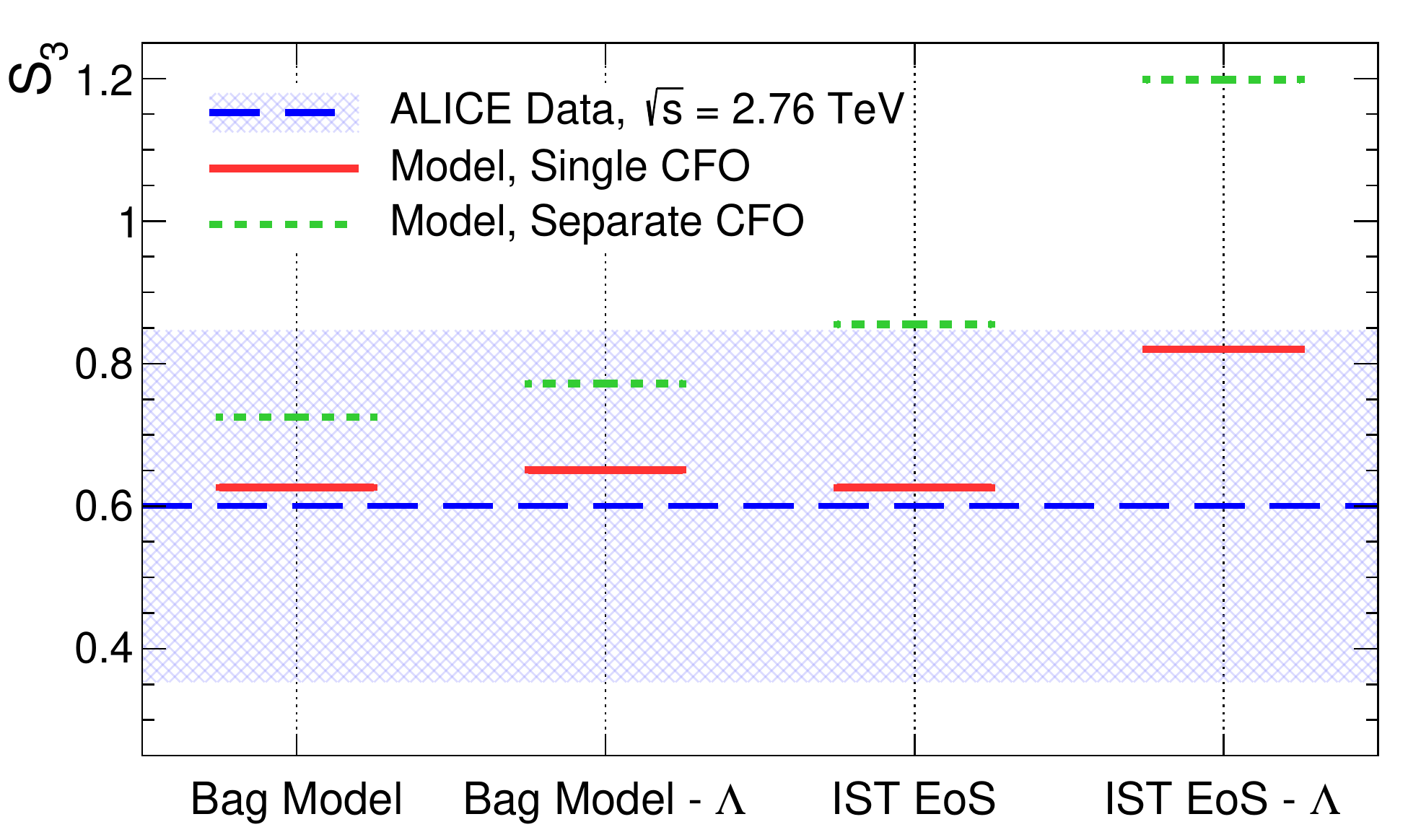}
}
	\caption{{\bf Left panel:}   $S_3$ ratio measured at $\sqrt{s_{NN}} = 200$ GeV by STAR  Collaboration vs. theoretical description obtained for  different CFO scenarios and EoS.
	{\bf Right panel:} The same as in the left panel, but for the data of  ALICE  Collaboration measured at   $\sqrt{s_{NN}} = 2.76$  TeV.
	}
	\label{KAB_Fig3}
\end{figure}
 
From Fig.~\ref{KAB_Fig1} one can see that for  the separate CFO of light nuclei IST$\Lambda$  EoS provides the CFO temperature of nuclei $T_A$ above 186 MeV (a similar result is found for IST EoS but with a larger  $\chi^2/dof$ value \cite{KAB_SepFO4}).  Note, however,  that according to the lattice version of  QCD  at vanishing values  of the  baryonic chemical potential  \cite{Bazavov:2018mes} it is rather problematic to use the hadronic EoS for such   CFO temperatures since this region is located  above the cross-over to the quark-gluon plasma. Although  for the separate  CFO scenario  all the  light nuclei data  are reproduced by the  IST$\Lambda$ EoS with  the deviation smaller than 1$\sigma$, this scenario can be ruled out by requiring  consistency with the lattice QCD results. The single CFO scenario of light nuclear clusters and hadrons corresponds to a CFO temperature  $T_A= T_h \simeq 168.30 \pm 3.85$ with $\chi^2_{tot}/dof \simeq  1.069$. 

As an independent benchmark  in favor of the single  CFO scenario for the STAR energies the  $S_3$ and $\overline{S}_3$ ratios 
\begin{equation}\label{eq_18}
	S_3 = \frac{{}^{3}_{\Lambda}{\rm H}}{{}^{3}{\rm He}} \times \frac{p}{\Lambda}, \quad  \overline{S}_3 = \frac{{}^{3}_{\overline{\Lambda}}\overline{{\rm H}}}{{}^{3}\overline{{\rm He}}} \times \frac{\overline{p}}{\overline{\Lambda}},
\end{equation}  
can be used \cite{KAB_SepFO4}. 
From the left panel of Fig.~\ref{KAB_Fig3} one can  see that the ratio $S_3$ provided by the STAR  Collaboration \cite{STARA1} is  accurately   reproduced only for the single CFO  scenario found by the IST$\Lambda$ EoS. It is remarkable that the data on the  $S_3$ and $\overline{S}_3$ ratios, which were not used in our fits, are reproduced by the most advanced  version of the HRGM for the single CFO scenario.   The quality of the light nuclei  STAR data description for this scenario is shown in the right panel of Fig.~\ref{KAB_Fig1}. 

From the analysis of the ALICE  $\sqrt{s_{NN}} = 2.76$ TeV  data  \cite{KAB_Ref1a,KAB_Ref1b,KAB_Ref1c} we obtained the opposite results.  In other words, the separate CFO scenario with $T_h \simeq 148.12 \pm 2.03$, $T_A \simeq 169.25 \pm 5.57$ and $\chi^2_{tot}/dof \simeq 0.753$ looks more preferable than the single CFO scenario with $T_h = T_A \simeq 150.29 \pm 1.92$ and $\chi^2_{tot}/dof \simeq 1.433$.  The details of   $\chi^2_{tot}/dof$ behavior and its parts  are shown in the left panel  of Fig.~\ref{KAB_Fig2}, while the right panel of this figure demonstrates the high quality of the nuclear data description achieved  by the IST$\Lambda$ EoS. In contrast to the STAR data,  the ALICE data on the $S_3$ ratio are inconclusive, since six  points out of eight ones found in our analysis  are located within the large error bars of  this quantity (see the right panel of Fig.~\ref{KAB_Fig3}). 

\section{Conclusions}

In this work we discussed a very  accurate description of the hadronic and light nuclear clusters data measured by the STAR Collaboration at  $\sqrt{s_{NN}} =200$ GeV and by  the ALICE LHC at $\sqrt{s_{NN}} =2.76$ TeV with the combined value  $\chi^2/dof  \simeq \frac{26.261}{18-3+17-3} \simeq 0.91$ for two best fits of  both data sets. Such a high quality of  data  description   is achieved by applying the new strategy of analyzing the light nuclear clusters data and by using the small value of the hard-core radius of the $\Lambda$-(anti-)hyperons $R_\Lambda=0.085$ fm found in  \cite{IST2,IST3} in the expressions for the classical second virial coefficients  of HTR and for the equivalent   hard-core radius of HTR.  

It is remarkable  that  the small value of  the hard-core radius of the $\Lambda$-(anti-)hyperons $R_\Lambda$  found in our previous works allowed us, for the first time, to  accurately describe the PHTR ratios measured by  the STAR  Collaboration. The  observed   high sensitivity of the HTR data to the classical hard-core radius of $\Lambda$-(anti-)hyperons allows us to hope  that in the future  one can  measure  the hard-core radii of other hyperons with high precision, if they form the hyper-nuclei.  

%%%%%%%%%%%%%%%%% Acknowledgements 
\ack 
{The work of  K.A.B., B.E.G, O.I.I., V.V.S., D.O.S. and G.M.Z.  was supported by the Program of Fundamental Research in High Energy and Nuclear Physics launched by the Section of Nuclear Physics of the National Academy of Sciences of Ukraine. 
V.V.S. and O.I.I. acknowledge the support  of   the Funda\c c\~ao para a Ci\^encia e Tecnologia (FCT), Portugal, under the project UID/04564/2020.
The work of   L.V.B., E.E.Z., O.V.V., E.S.Z. and K.A.B. was supported in part  by the  projects  RFBR 18-02-484, RFBR 18-02-485, CPEA-LT-2016/10094, UTF-2016/10076 and  NFR-255253/F50. 
D.B.B. is thankful to the RFBR for the partial support  under the grant No. 18-02-40137.
A.V.T.  acknowledges the partial support from the RFBR under the grant No. 18-02-40086 and from the Ministry of Science and Higher Education of the Russian Federation,  Project ``Fundamental properties of elementary particles and cosmology" No
0723-2020-0041.
The authors are grateful to the COST Action CA15213 ``THOR``  for supporting their networking.}

%\medskip

\end{document}